\documentclass[%
 reprint,
 amsmath,amssymb,
 aps,
]{revtex4-2}
\usepackage{graphicx}
\usepackage{dcolumn}
\usepackage{bm}
\usepackage{multirow}

\usepackage{color}

\usepackage[colorlinks,citecolor=blue]{hyperref}

\usepackage{float}
\usepackage{rotating}
\usepackage{upgreek}
\usepackage{subcaption}
\usepackage[normalem]{ulem} 

\usepackage{comment}
\newcommand{\numu}{$\nu_{\mu}$}
\newcommand{\nue}{$\nu_{e}$}
\newcommand{\nueb}{$\bar{\nu}_{e}$}
\newcommand{\thetamu}{$\theta_{23}$}
\newcommand{\thetamubar}{$\overline{\theta}_{23}$}
\newcommand{\numub}{$\overline{\nu}_{\mu}$}
\newcommand{\sinsqthetamu}{$\sin^2\theta_{23}$}
\newcommand{\sinsqthetamubar}{$\sin^2\overline{\theta}_{23}$}
\newcommand{\sinsqtwothetamu}{$\sin^22\theta_{23}$}
\newcommand{\sinsqtwothetamubar}{$\sin^22\overline{\theta}_{23}$}
\newcommand{\dmatm}{$\Delta m^2_{31}$}
\newcommand{\dmbaratm}{$\Delta \overline{m}^2_{31}$}

\newcommand{\dcp}{$\delta_{CP}$}
\newcommand{\dcptdm}{$\delta_{\nu\overline{\nu}}(\Delta m^2_{31})$}
\newcommand{\dcpttheta}{$\delta_{\nu\overline{\nu}}(\sin^2\theta_{23})$}

\newcommand{\nova}{NO$\nu$A}
\newcommand{\numode}{$\nu$-mode}
\newcommand{\antinumode}{$\bar{\nu}$-mode}

\begin{document}


\title{Stringent constraint on CPT violation with the synergy of T2K-II, NO$\nu$A extension, and JUNO}

\author{T. V. Ngoc$^{1,2}$}
\thanks{Corresponding Author tranngocapc06@ifirse.icise.vn}

\author{S. Cao$^{1}$}%
\author{N. T. Hong Van$^{3}$}
\author{P. T. Quyen$^{1}$}
\affiliation{\vspace*{2mm}
$^1$\textit{Institute For Interdisciplinary Research in Science and Education},\\ 
\it{ICISE, Quy Nhon, Vietnam.\vspace*{2mm}\\}
$^2$\textit{Graduate University of Science and Technology, Vietnam Academy of Science and Technology, Hanoi, Vietnam.}\vspace*{2mm}\\
$^3$\textit{Institute of Physics, Vietnam Academy of Science and Technology, Hanoi, Vietnam}.\\
}

\date{\today}

\begin{abstract}
Neutrino oscillation experiments have measured precisely at few percent levels the mass-squared differences ($\Delta m^2_{21}$, $\Delta m^2_{31}$) of three neutrino mass eigenstates, and the three leptonic mixing angles ($\theta_{12},~ \theta_{13},~ \theta_{23}$) by utilizing both neutrino and anti-neutrino oscillations. The possible CPT violation may manifest itself in the difference of neutrino and anti-neutrino oscillation parameters, making these experiments promising tools for testing CPT invariance at unprecedented precision. We investigate empirically the sensitivity of the CPT test via the difference in mass-squared splittings ($\Delta m^2_{31} - \Delta \overline{m}^2_{31}$) and in leptonic mixing angles ($\sin^2\theta_{23} - \sin^2\overline{\theta}_{23}$) with the synergy of T2K-II, NO$\nu$A extension, and JUNO experiments. If the CPT symmetry is found to be conserved, the joint analysis of the three experiments will be able to establish limits of $|\Delta m^2_{31} - \Delta \overline{m}^2_{31}|<5.3\times 10^{-3} \text{eV}^2$ and $|\sin^2\theta_{23} - \sin^2\overline{\theta}_{23}|<0.10$ at 3$\sigma$ confidence level (C. L.) on the possible CPT violation, extending substantially the current bound of these parameters. We find that with ($\Delta m^2_{31} - \Delta \overline{m}^2_{31}$), the dependence of the statistical significance on the relevant parameters to exclude the CPT conservation is marginal, and that, if the difference in the best-fit values of $\Delta m^2_{31}$ and $\Delta \overline{m}^2_{31}$ measured by MINOS(+) and NO$\nu$A persists as the \textit{true}, the combined analysis will rule out the CPT conservation at 4$\sigma$ C. L.. With the ($\sin^2\theta_{23} - \sin^2\overline{\theta}_{23}$), the statistical significance to exclude CPT invariance depends strongly on the \textit{true} value of $\theta_{23}(\overline{\theta}_{23})$ mixing angle. In case of maximal mixing of $\theta_{23}$, as indicated by the current T2K and NO$\nu$A measurements, the CPT conservation will be excluded at 3$\sigma$ C. L. or higher if the difference in the best-fit values of $\theta_{23}$ and $\overline{\theta}_{23}$ remains as the \textit{true}.
\end{abstract}

\maketitle


\section{\label{intro}CPT test with neutrino}
The CPT theorem, which connects three discrete symmetries: charge conjugation (C), parity (P), and time reversal (T), and has been theoretically proved in different ways~\cite{Luders1954,1955nbdpbookPauli,Jost1957,Schwinger1958,johnbell}, states that any Lorentz invariant local quantum field theory of point-particle must be CPT invariant. If it is discovered that CPT symmetry is not conserved, one of the three foundational assumptions (Lorentz invariance, Hamiltonian hermiticity, and locality) must be sternly reconsidered. A consequence of the CPT invariance is that the particle and its anti-particle must have the same energy spectra. This important property opens a possibility for direct testing CPT invariance by comparing the mass spectra, or other properties such as lifetime or magnetic moment of a particle and its anti-particle. Ref.~\cite{Kosteleck_2011} provides the latest results on Lorentz and CPT violation searches in the context of Standard Model Extension. A summary of the model-independent CPT testing based on different properties of the different systems of particle and anti-particle can be found in Ref.~\cite{CPTsummary2022}. In terms of relative precision, the most stringent constraint on the CPT test was achieved on the neutral kaon system \cite{pdg2020} 
\begin{equation}\label{eq:kaoncpt}
\left|\frac{m(K^\circ) - m(\overline{K}^\circ)}{m_K}\right| < 6\times 10^{-19} ~\text{at}~ 90\% ~\text{C. L.}
\end{equation}
As pointed out in Ref.~\cite{Murayama_2004}, when expressed in terms of the mass-squared difference, the bound on the $K^\circ - \overline{K}^\circ$ mass difference does not appear to be formidable. From  Eq.(\ref{eq:kaoncpt}), one can get  
\begin{equation}\label{eq:kaoncpt_msq}
    |m^2(K^\circ)- m^2(\overline{K}^\circ)| < 0.3 ~\text{eV}^2.
\end{equation}
Comparing this to the two mass-squared differences of the three neutrino mass eigenstates~\cite{pdg2020}, $m_{\nu2}^2-m_{\nu1}^2\approx 7.4\times 10^{-5}\ \text{eV}^2$ and $|m_{\nu3}^2-m_{\nu2}^2|\approx 2.45\times 10^{-3}\ \text{eV}^2$, it becomes clear that neutrino measurements, rather than neutral kaons, provide the best constraint on the CPT test in terms of the mass-squared difference~\cite{Barger_2000,Murayama_2004}. The aforementioned neutrino mass-squared differences come from measuring the neutrino oscillation, which is a macroscopic quantum phenomenon establishing that neutrinos are massive and thus beyond the Standard Model's description. It is worth noting that the neutrino mass spectrum cannot be calculated solely from neutrino oscillations, but must be combined with cosmological constraints and beta decay, as recently discussed in Ref.~\cite{Cao:2021ptu}. Neutrinos, unlike neutral B and K mesons, are neutral elementary particles, and it is intriguing that this particle could be a Majorana particle, where neutrino and anti-neutrino are indistinguishable in the conventional sense of the CPT invariant paradigm. The neutrino nature under the CPT-violating scenario has been explored in Ref.~\cite{Barenboim:2002hx}. Here we focus on the phenomenological consequence of the CPT violation in the observable neutrino oscillation.

In context of three-flavor PMNS framework~\cite{maki1962remarks,pontecorvo1958mesonium}, for a given propagation distance $L$ and matter density $\rho$, the probabilities $(P_{\nu_{\alpha}\rightarrow\nu_{\beta}},\ P_{\overline{\nu}_{\alpha}\rightarrow \overline{\nu}_{\beta}} )$ for a neutrino and anti-neutrino at a specific energy $(E_{\nu},E_{\overline{\nu}})$ oscillating from one flavor $(\nu_{\alpha},\overline{\nu}_{\alpha})$ to another flavor $(\nu_{\beta},\overline{\nu}_{\beta})$ are completely and commonly described with six oscillation parameters including three leptonic mixing angles $(\theta_{12},\ \theta_{13},\ \theta_{23})$, one Dirac \textit{CP}-violating phase \dcp, and two mass-squared differences $(\Delta m^2_{21},\ \Delta m^2_{31}$). Under CPT symmetry, the neutrino and anti-neutrino oscillation probabilities are well connected as follows:
\begin{align*}
    \label{eq:cptholdprob}\nonumber
    P_{\nu_\alpha \rightarrow \nu_\beta} \xrightarrow{\text{CPT}} &P_{\overline{\nu}_\beta \rightarrow \overline{\nu}_\alpha}=P_{\nu_\alpha \rightarrow \nu_\beta} \\
    &=f(\theta_{12},~ \theta_{13},~ \theta_{23},~\delta_{\text{CP}},~\Delta m^2_{21},~\Delta m^2_{31}). 
\end{align*}

If the CPT is violated in neutrino sector, the underlying sets of oscillation parameters in neutrino and anti-neutrino may differ. Empirically, we assume
\begin{equation}\label{eq:nuparas}
     P_{\nu_\alpha \rightarrow \nu_\beta} = f(\theta_{12},~ \theta_{13},~ \theta_{23},~ \delta_{\text{CP}},~  \Delta m^2_{21},~ \Delta m^2_{31}),
\end{equation}
for describing the neutrino oscillations, and 
\begin{equation}\label{eq:nubarparas}
    P_{\overline{\nu}_\beta \rightarrow \overline{\nu}_\alpha} = f(\overline{\theta}_{12},~ \overline{\theta}_{13},~ \overline{\theta}_{23},~ \overline{\delta}_{\text{CP}},~  \Delta \overline{m}^2_{21},~ \Delta \overline{m}^2_{31}),
\end{equation}
for anti-neutrino oscillations. 

If there are observable differences in the parameters of the two sets, it may indicate a CPT violation in the lepton sector. 
Since the discovery of neutrino oscillations~\cite{RevModPhys.88.030501,RevModPhys.88.030502} at the end of the twentieth century, neutrino oscillation experiments~\cite{pdg2020} using both natural and man-made neutrino sources have transitioned into the precision measurement phase of three mixing angles and two mass-squared differences, and being explored three remained known unknowns including the neutrino mass ordering, whether CP is violated, and whether the mixing angle $\theta_{23}$ is maximal ($\theta_{23}=45^{\circ}$) or belong to a lower ($\theta_{23}<45^{\circ}$) or higher ($\theta_{23}>45^{\circ}$) octant. Each experiment is typically sensitive to a subset of the oscillation parameters but not the entire set. The experiments with solar neutrinos provide the most constraints on the $(\theta_{12},~\Delta m^2_{21})$ parameters while the reactor-based long-baseline neutrino (R-LBL) experiments can measure precisely the ($\overline{\theta}_{12},~\Delta {\overline{m}}^2_{21}$) parameters. The reactor-based short-baseline (order of 1~km) neutrino (R-SBL) experiments play a central role in measuring the ($\overline{\theta}_{13},~\Delta \overline{m}^2_{31}$) parameters. The under-developing reactor-based medium-baseline neutrino (R-MBL) experiment JUNO, which will be discussed later, takes advantage of interference of oscillations at different wavelengths, huge statistics, and good energy resolution to achieve sub-percent precision in measuring the ($\overline{\theta}_{12},~\Delta {\overline{m}}^2_{21},~\Delta \overline{m}^2_{31}$) parameters. Experiments with the atmospheric neutrino and accelerator-based neutrino sources can precisely measure the ($\theta_{23},~\overline{\theta}_{23},~\Delta m^2_{31},~\Delta \overline{m}^2_{31}$) parameters. Besides, this type of experiment is also sensitive to the $(\theta_{13},~\overline{\theta}_{13})$ parameters, but the precision of these parameters is much lower in comparison to the R-SBL experiment due to the statistical limit and their strong correlation with two known unknowns, \textit{CP}-violating phase and neutrino mass ordering. Although there is some hint~\cite{T2Knature2020} of non-zero \textit{CP}-violating phase \dcp, precise measurement on this parameter is not possible until the next generation of the accelerator-based long-baseline (A-LBL) experiments. It is provided in Ref.~\cite{Barenboim2020} the most recent update at $3\sigma$ confidence level (C. L.) on the bounds of CPT violation on each individual parameter with global neutrino data.
\begin{eqnarray}\label{eq:nucptbound} \nonumber
|\delta_{\nu\overline{\nu}}(\Delta m^2_{21})| &<& 4.7\times 10^{-5}~\text{eV}^2, \\ \nonumber
|\delta_{\nu\overline{\nu}}(\Delta m^2_{31})| &<& 2.5\times 10^{-4}~\text{eV}^2,\\ \nonumber
|\delta_{\nu\overline{\nu}}(\sin^2\theta_{12})| &<& 0.14, \\ \nonumber
|\delta_{\nu\overline{\nu}}(\sin^2\theta_{13})| &<& 0.029, \\ 
|\delta_{\nu\overline{\nu}}(\sin^2\theta_{23})| &<& 0.19, 
\end{eqnarray}
where $\delta_{\nu\overline{\nu}}(X) = X - \overline{X}$ for the X neutrino oscillation parameter and the $\overline{X}$ anti-neutrino oscillation parameter.
In this study, we focus on the synergy between two on-going A-LBL experiments (T2K and \nova) and one under-developing R-MBL experiment (JUNO) to explore the potential sensitivity to the measurement of \dcptdm\ and \dcpttheta\ parameters. The A-LBL experiments utilize the highly intense beam of the almost pure muon neutrinos $\nu_{\mu}$ and muon anti-neutrinos $\overline{\nu}_{\mu}$ for measuring the four transitions categorized into two channels, \textit{appearance} channels $(\nu_\mu \rightarrow \nu_e, ~\overline{\nu}_\mu \rightarrow \overline{\nu}_e)$, and \textit{disappearance} channels $(\nu_\mu \rightarrow \nu_\mu, ~\overline{\nu}_\mu \rightarrow \overline{\nu}_\mu)$.  
While the \textit{appearance} channels are sensitive to a wider subset of parameters and being explored for searching the CP violation in the lepton sector, measuring $(\nu_\mu \rightarrow \nu_e, ~\overline{\nu}_\mu \rightarrow \overline{\nu}_e)$ is not sufficient to test CPT directly since the corresponding CPT-mirrored processes are missing. The  \textit{disappearance} channels, on the other hand, are well-suited for testing CPT since they are two CPT-mirrored processes. We characterize the difference in the probabilities of the muon neutrino \textit{disappearance} and muon anti-neutrino \textit{disappearance}, $\mathcal{A}^{\text{CPT}}_{\mu\mu} = P_{\nu_\mu \rightarrow \nu_\mu} - P_{\overline{\nu}_\mu \rightarrow \overline{\nu}_\mu}$ as an observable measure of the CPT-violating effect. 

\begin{figure*}
    \includegraphics[scale=0.8]{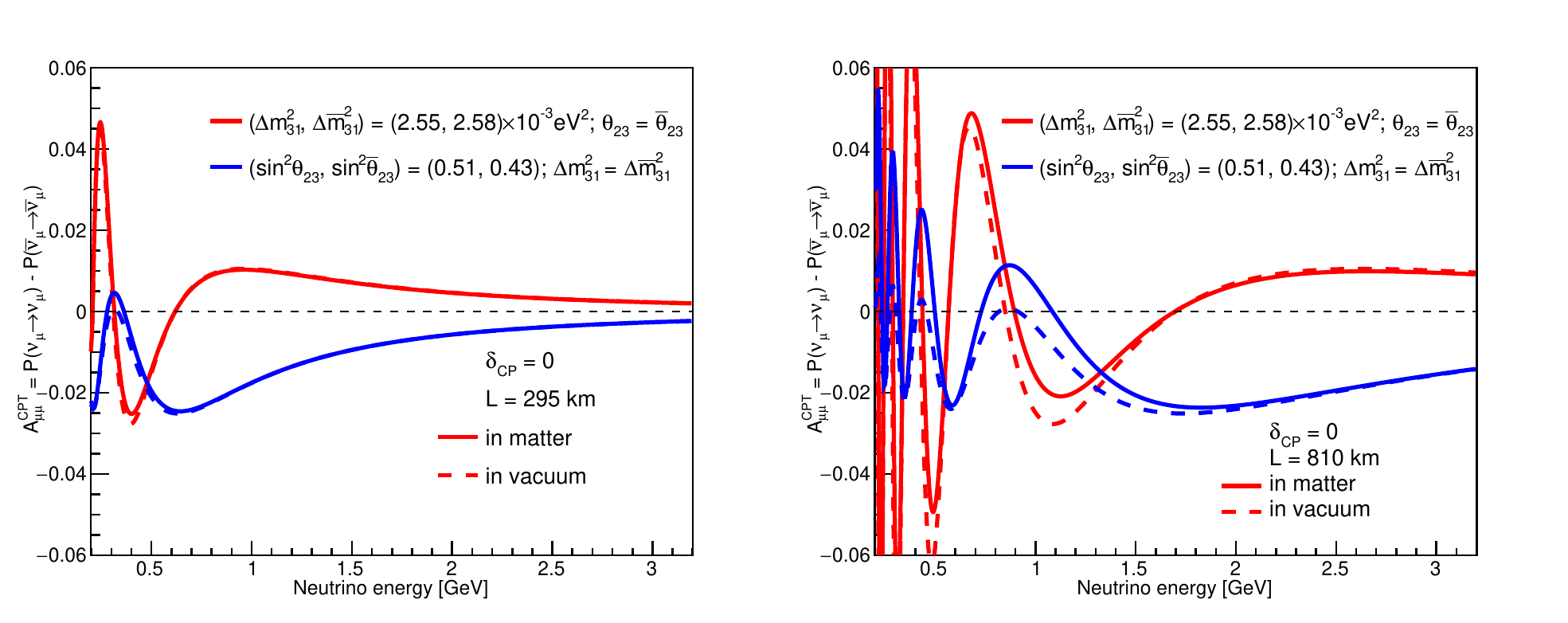}
    \caption{\label{fig:acpt_t2k} CPT asymmetries in  \textit{disappearance} channels for T2K baseline L = 295 km (left) and NO$\nu$A baseline L = 810 km (right). The differences in solid lines and dashed lines indicate extrinsic CPT effects caused by matter.}
\end{figure*}

The observable asymmetry $\mathcal{A}^{\text{CPT}}_{\mu\mu}$ may consist of two parts: \textit{intrinsic} CPT asymmetry and \textit{extrinsic} CPT asymmetry caused by differences in interactions between neutrinos and anti-neutrinos with the matter of the propagation medium \cite{Barenboim_2001,Xing_2002,Bernabeu_2002,Jacobson_2004,OHLSSON2015482}. Fig.~\ref{fig:acpt_t2k} illustrates the CPT asymmetries $\mathcal{A}^{\text{CPT}}_{\mu\mu}$  calculated in vacuum and in the matter presence at baselines of the T2K experiment (L = 295~km) and of the \nova\ experiment (L = 810~km). Here we take the best-fit values of the mainly involved $(\Delta m^2_{31},~\Delta \overline{m}^2_{31},~\theta_{23},~ \overline{\theta}_{23})$ parameters from the recent T2K results~\cite{abe2021t2k} and of the others from the global data analysis~\cite{deSalas2020}, which are summarized in Table~\ref{tab:nuoscpara}. It is worthy to notice that our work with the muon-neutrino and muon-anti-neutrino disappearance data sample is insignificantly affected by the uncertainty in our understanding of $(\theta_{12},\ \overline{\theta}_{12},\ \theta_{13},\ \overline{\theta}_{13},\ \delta_{\text{CP}}, \ \overline{\delta}_{\text{CP}},\ \Delta m^{2}_{21}, \Delta \overline{m}^{2}_{21})$ parameters. The primary driving parameters in this study are $(\theta_{23},\ \overline{\theta}_{23},\ \Delta m^{2}_{31},\ \Delta \overline{m}^{2}_{31})$ parameters.
\begin{table}
    \centering
    \begin{tabular}{l|c}
    \hline\hline
    Parameter & Value\\\hline
    $\sin^{2}\theta_{23}$ & $0.51$\\
    $\sin^{2}\overline{\theta}_{23}$ & $0.43$\\
    $\Delta m^{2}_{31} $ & $2.55\times 10^{-3}\text{eV}^{2}$\\
    $\Delta \overline{m}^{2}_{31}$ & $2.58\times 10^{-3}\text{eV}^{2}$\\
    & \\        
    $\sin^{2}\theta_{12},~ \sin^{2}\overline{\theta}_{12}$ & $0.318$\\    $\sin^{2}\theta_{13},~ \sin^{2}\overline{\theta}_{13}$ & $0.022$\\ $\delta_{\text{CP}},~ \overline{\delta}_{\text{CP}}$ & $1.08\pi$~ rad\\
    $\Delta m^{2}_{21},~ \Delta \overline{m}^{2}_{21}$ & $7.50\times10^{-5}\text{eV}^{2}$\\
    \hline
    \end{tabular}
    \caption{Values of nominal parameters, taken from the recent T2K measurements~\cite{abe2021t2k} of muon-neutrino and muon-anti-neutrino \textit{disappearances} and from the global analysis of the neutrino oscillation data~\cite{deSalas2020}. Our work utilising the data samples of muon-neutrino and muon-anti-neutrino disappearance is insignificantly affected by uncertainty of ($\theta_{12}$, $\ \overline{\theta}_{12}$, $\theta_{13}$, $\overline{\theta}_{13}$, $\delta_{\text{CP}}$,  $\overline{\delta}_{\text{CP}}$, $\Delta m^{2}_{21}$, $\Delta \overline{m}^{2}_{21}$) parameters.}
    \label{tab:nuoscpara}
\end{table}

By comparing the in-vacuum and in-matter cases, it shows that the matter effect with \nova\ is more visible than T2K due to the longer baseline. However, for both cases, at the peak of experimental neutrino spectra (0.6~GeV for T2K and 2.0~GeV for \nova), the matter effect is marginal and thus the CPT test is relatively transparent. In addition, it is observed that 1\% difference between the two mass-square splittings translates to approximately 1\% difference in CPT asymmetry $\mathcal{A}^{\text{CPT}}_{\mu\mu}$. Regarding to  $(\theta_{23},~ \overline{\theta}_{23})$-dependence, since $P_{\nu_\mu \rightarrow \nu_\mu}$ and $ P_{\overline{\nu}_\mu \rightarrow \overline{\nu}_\mu}$ up to the first order of approximation, are proportional to $\sin^22\theta_{23}$ and $\sin^22\overline{\theta}_{23}$, respectively. In Fig.~\ref{fig:acpt_t2k}, about 15\% difference between \sinsqthetamu\ and \sinsqthetamubar\ is converted to 2\% difference between $\sin^22\theta_{23}$ and $\sin^22\overline{\theta}_{23}$ and results in around 2\% of the CPT asymmetry. Furthermore, it is worth noting that the possible CPT asymmetry, if happened, between the $\nu_{\mu}$ and $\overline{\nu}_{\mu}$\textit{disappearances} does not depend on the Dirac \textit{CP}-violating phase. This channel, on its own, is less sensitive to neutrino mass ordering. The scenario, where neutrino follows the \textit{normal} ordering while the anti-neutrino follows the \textit{inverted} ordering or vice versa, implies the CPT violation. This scenario can be tested by comparing the neutrino mass ordering measured with neutrino-mode data samples in the A-LBL experiment to the measurement from the JUNO experiment. The work in Ref.~\cite{SonCao2020} shows that under the CPT-invariant assumption, combining these three experiments will resolve the neutrino mass ordering completely. It will be exciting, however, if the A-LBL experiments and JUNO point separately to different mass orders with high statistical significance. In this study, we assume that neutrino and anti-neutrino masses are ordered similarly. 

Table~\ref{tab:atmos_paras_experiments} summarize the measurements of the $(\Delta m^2_{31},~\Delta \overline{m}^2_{31},~\theta_{23},~ \overline{\theta}_{23})$ parameters with the first generation of the A-LBL experiment MINOS~\cite{Adamson2013,minos2020}, on-going second generation T2K~\cite{abe2021t2k}, \nova\ \cite{nova20192d118E20POT}, and precise constraint of the \dmbaratm\ parameter from the R-SBL experiment Daya Bay~\cite{dayabay2022}.
\begin{table}
    \centering
    \begin{tabular}{c|c|c|c|c}
    \hline
    \hline
         & MINOS(+) & T2K & \nova & Daya Bay \\
    \hline
      $\Delta m^2_{31}/10^{-3} \text{eV}^2$  & $2.48^{+0.08}_{-0.09}$ & $2.55^{+0.08}_{-0.09}$ & $2.56^{+0.07}_{-0.09}$ & -  \\
      $\Delta\overline{m}^2_{31}/10^{-3} \text{eV}^2$ & $2.55^{+0.23}_{-0.25}$  & $2.58^{+0.18}_{-0.13}$ & $2.63^{+0.12}_{-0.13}$ & $2.53^{+0.06}_{-0.06}$ \\
      \sinsqthetamu & $0.43_{-0.04}^{+0.20}$ & $0.51^{+0.06}_{-0.07}$ & $0.51^{+0.06}_{-0.06}$ & -  \\
      \sinsqthetamubar  & $0.41^{+0.05}_{-0.08}$ & $0.43^{+0.21}_{-0.05}$ & $0.41^{+0.04}_{-0.03}$ & -  \\
      \hline
    \end{tabular}
    \caption{Measurements of the $(\Delta m^2_{31},~\Delta \overline{m}^2_{31},~\theta_{23},~ \overline{\theta}_{23})$ parameters, which govern the muon neutrino and muon anti-neutrino \textit{disappearances}, from different experiments: MINOS(+) \cite{Adamson2013,minos2020}, T2K~\cite{abe2021t2k}, \nova\ \cite{nova20192d118E20POT}, Daya Bay~\cite{dayabay2022}. \textit{Normal} neutrino mass ordering is assumed.}
    \label{tab:atmos_paras_experiments}
\end{table}
It is shown that \dmatm\ is measured with about 3\% precision with the A-LBL experiments, while \dmbaratm\ is measured with about 10\% precision, which can be complemented with the R-SBL experiment with 2.3\% precision. For the mixing angle, the precision is varied among experiments due to the fact that we are unsure whether $(\sin^2\theta_{23}, \sin^2\overline{\theta}_{23})$ is maximal or belong to a specific octant. The neutrino and anti-neutrino involved parameters agree within 1$\sigma$ C. L.

In this paper, we will investigate the prospects of testing the possible CPT violation via the applicably sensitive \dcptdm\ and \dcpttheta\ parameters with the synergy of T2K-II, \nova\ extension (for convenience, we will denote it \nova-II from now on), and JUNO experiments. In particular, we focus on the use of data samples of the $\nu_{\mu}$ and $\overline{\nu}_{\mu}$\textit{disappearance} channels from the T2K-II and \nova-II experiments in combination with the \textit{disappearance} of $\overline{\nu}_e$ collected by the JUNO experiment before 2028, where we expect the operational start of the next generation A-LBL experiments. The paper is organized as follows. We describe the simulation of T2K-II, \nova-II and JUNO experiments in Sec.~\ref{sec:simulate}. The possibly established bounds of the manifested quantities \dcptdm\ and \dcpttheta\ of the CPT violation are presented in Sec.~\ref{sec:CPTbound}. Further investigation into the potential significance of CPT-invariant exclusion and its robustness against the variation of the underlying physical parameters are discussed in Sec.~\ref{sec:sensiCPT}. Finally, we conclude our study in Sec.~\ref{sec:fin}.

\section{\label{sec:simulate} Experimental simulation}
The General Long Baseline Experiment Simulator (GLoBES)~\cite{Huber:2004ka,huber2007new} is a sophisticated but flexible framework to simulate, explore the physic potentials of neutrino experiments and fit the experimental data. By default, GLoBES assumes that the oscillation parameters for neutrinos and anti-neutrinos in Eq.~(\ref{eq:nuparas}) and Eq.~(\ref{eq:nubarparas}) are identical or CPT-invariant. We extend the package to describe the neutrino and anti-neutrino oscillations independently.
 
For the oscillation probability formula, we follow the analytical expressions in Ref.~\cite{Barger1980}. Neutrino (anti-neutrino) oscillation in matter depends on nine variables, including six oscillation parameters listed in Eq.~(\ref{eq:nuparas}) for neutrino (or Eq.~(\ref{eq:nubarparas}) for anti-neutrino), as well as neutrino energy $E_{\nu}$, the propagation distance $L$, and the matter density $\rho$. For the CPT test, the oscillation parameters of neutrinos and anti-neutrinos can be treated independently, thus having twelve oscillation parameters as a complete set. However, for this particular study, since the A-LBL experiments have no sensitivity to the solar parameter, we keep $\theta_{12} = \overline{\theta}_{12};\ \Delta m^2_{21}= \Delta \overline{m}^2_{21};\ \theta_{13} = \overline{\theta}_{13};\ \delta_{\text{CP}}=\overline{\delta}_{\text{CP}}$ and fixed practically. Four independent parameters of interests $(\Delta m^2_{31},\ \Delta \overline{m}^2_{31},\ \theta_{23}, \ \overline{\theta}_{23})$ remains.

T2K~\cite{Abe:2011ks} and \nova ~\cite{ayres2007nova} are two world-leading A-LBL experiments. For convenience, we denote T2K run up to 2027 by T2K-II and \nova\ extension up to 2024 by \nova-II. The similarity in experimental configuration and operating principle makes it interesting to have a joint fit between the two experiments~\cite{christophe_bronner_2022_6683821,jeff_hartnell_2022_6683827}. Both experiments use intense muon (anti-)neutrino beams created by accelerators to study oscillation phenomena. The off-axis technique adopted by both experiments can produce a narrow-band beam of neutrinos to enhance the sensitivity of neutrino oscillation measurements and mitigate the effect of possible bias in the neutrino energy reconstruction from their interaction products. The ability to focus either positive or negative particles (mainly pions and kaons) offers the A-LBL experiment a unique opportunity to operate in both neutrino-mode and anti-neutrino-mode. This important feature enables the testing of CPT invariance in the A-LBL experiments. JUNO~\cite{djurcic2015juno} is a R-MBL experiment which studies electron anti-neutrino \textit{disappearance} ($\overline{\nu}_e \rightarrow \overline{\nu}_e$). The experiment uses electron anti-neutrino flux produced from nuclear reactors to study neutrino oscillation at a medium baseline (about 50~km) to take advantage of the interference of two oscillation lengths, which are driven by two mass-squared splittings, $\Delta \overline{m}^2_{21}$ and \dmbaratm. Achieving a neutrino energy resolution of less than $3\%$ is essential for JUNO to resolve these two oscillation patterns and measure oscillation parameters $\sin^2\overline{\theta}_{12},~\Delta \overline{m}^2_{21}$ and $|\Delta \overline{m}^2_{31}|$ at precision less than 0.5\%~\cite{juno_2022}. The JUNO experiment also can resolve neutrino mass hierarchy at $3\sigma$ C. L. after six years of operation. Combining data samples from JUNO and from the A-LBL experiments, T2K-II and \nova-II, will definitely resolve the neutrino mass ordering~\cite{SonCao2020}.

We follow closely the experimental specifications for T2K-II, \nova-II, and JUNO in the Ref.~\cite{SonCao2020}, except for some updates in T2K-II and JUNO. In original proposal~\cite{abe2016sensitivity}, T2K-II is expected to operate until 2027, exposing $20\times 10^{21}$ protons-on-target (POT). According to the most recent plan~\cite{christophe_bronner_2022_6683821}, statistics may be cut in half. Thus, we use $10\times 10^{21}$~POT for T2K-II in this work. We also updated the T2K flux, which was released in 2020 \cite{t2k_collaboration_2021_flux}. For JUNO, a total thermal of 26.6~GWth \cite{juno_2022} is used instead of 36~GWth as in the previous report. Also, the energy resolution is set at 2.9$\%$ \cite{juno_2022} to reflect closely the JUNO's prospect.

In terms of the data samples for analysis, for T2K-II and \nova-II, we used the \textit{disappearance} channels only, with statistics equally divided into \numode\ and \antinumode. As we will show later in Sec.~\ref{sec:CPTbound}, the CPT test on the \dcptdm\ will be limited due to the precision of \dmatm\ measurement by the A-LBL experiment, and thus the bound established in this parameter can be elevated if we have more neutrino data. However, this scenario is unlikely since running an experiment in anti-neutrino mode is very important for the CP violation measurement. For JUNO, $\overline{\nu}_e$ \textit{disappearance} data is used. 
We assume neutrino masses are in \textit{normal} ordering throughout the study in Sec.~\ref{sec:CPTbound} and Sec.~\ref{sec:sensiCPT}. The study in Sec.~\ref{sec:CPTbound} is done with the values of nominal parameters listed in Table \ref{tab:nuoscpara}, in which we follow the measurements of T2K \cite{abe2021t2k} for atmospheric parameters ($\Delta m^2_{31},\ \Delta \overline{m}^2_{31},\ \sin^2\theta_{23},\ \sin^2\overline{\theta}_{23}$) and global fit \cite{deSalas2020} for the rest. 

The bounds and the sensitivities to rule out CPT invariant hypothesis with $\delta_{\nu\overline{\nu}}(X)$ parameter are explored. The $\chi^2$ of individual experiment is calculated for given \textit{true} values of $X$ for neutrinos and $\overline{X}$ for anti-neutrinos, where $X$ can be \sinsqthetamu\ or \dmatm. We use a log-likelihood $\chi^2$ function for T2K-II and \nova-II, while a Gaussian formula is used for JUNO due to its high statistics. The calculation of $\chi^2$ is then minimized over the nuisance parameters and other oscillation parameters except for $X$ and $\overline{X}$. The two-dimensional distributions of $\Delta \chi^2$ which is the sum of all the individual ones of the three experiments, are obtained. The minimum of $\Delta \chi^2$ as a function of $\delta_{\nu\overline{\nu}} (X) = X - \overline{X}$ is then found. The statistical significance of excluding CPT conservation is expressed as the squared root of the minimum $\Delta\chi^2$. 

\section{\label{sec:CPTbound}Possibly established bounds of \dcptdm\ and \dcpttheta\ on CPT violation}
In this study, assuming that CPT is exactly conserved or extremely small for detection, we estimate the expected bound of the two sensitive parameters, asymmetry in the mass-squared differences \dcptdm\ and asymmetry in the leptonic mixing angles \dcpttheta, on the possible CPT violation. In particular, $\Delta {m}^2_{31}=\Delta\overline{m}^2_{31} = 2.55\times 10^{-3}\ \text{eV}^2$ and $\sin^2\theta_{23}=\sin^2\overline{\theta}_{23}=0.51$, which are the T2K's best-fit points with recent measurement~\cite{abe2021t2k}, are assumed to be \textit{true}. To compute the allowed region of the \dcptdm\ and \dcpttheta\ parameters, we build up the $\chi^2$ profiles on a two-dimensional grid points of neutrino and anti-neutrino corresponding parameters ($\Delta {m}^2_{31},\ \Delta\overline{m}^2_{31}$) and  ($\sin^2\theta_{23},\ \sin^2\overline{\theta}_{23}$), respectively. The $\chi^2$ profiles take into account the correlations among the oscillation parameters. The $\Delta \chi^2$ profiles are attained by subtracting to the minimum value of the according $\chi^2$, which is essentially located at the \textit{true} values.
\begin{figure*}
\includegraphics[width=0.45\textwidth]{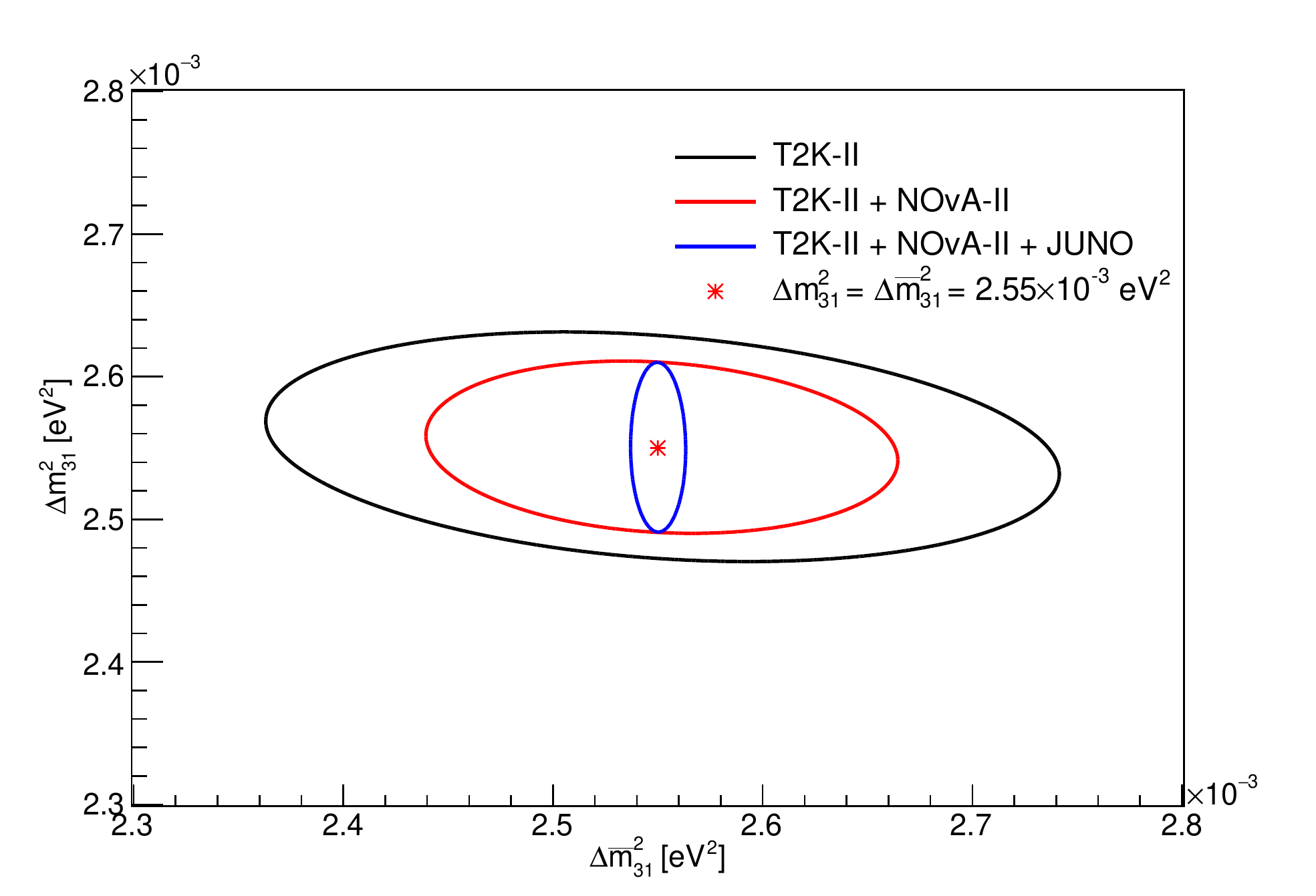}
\includegraphics[width=0.45\textwidth]{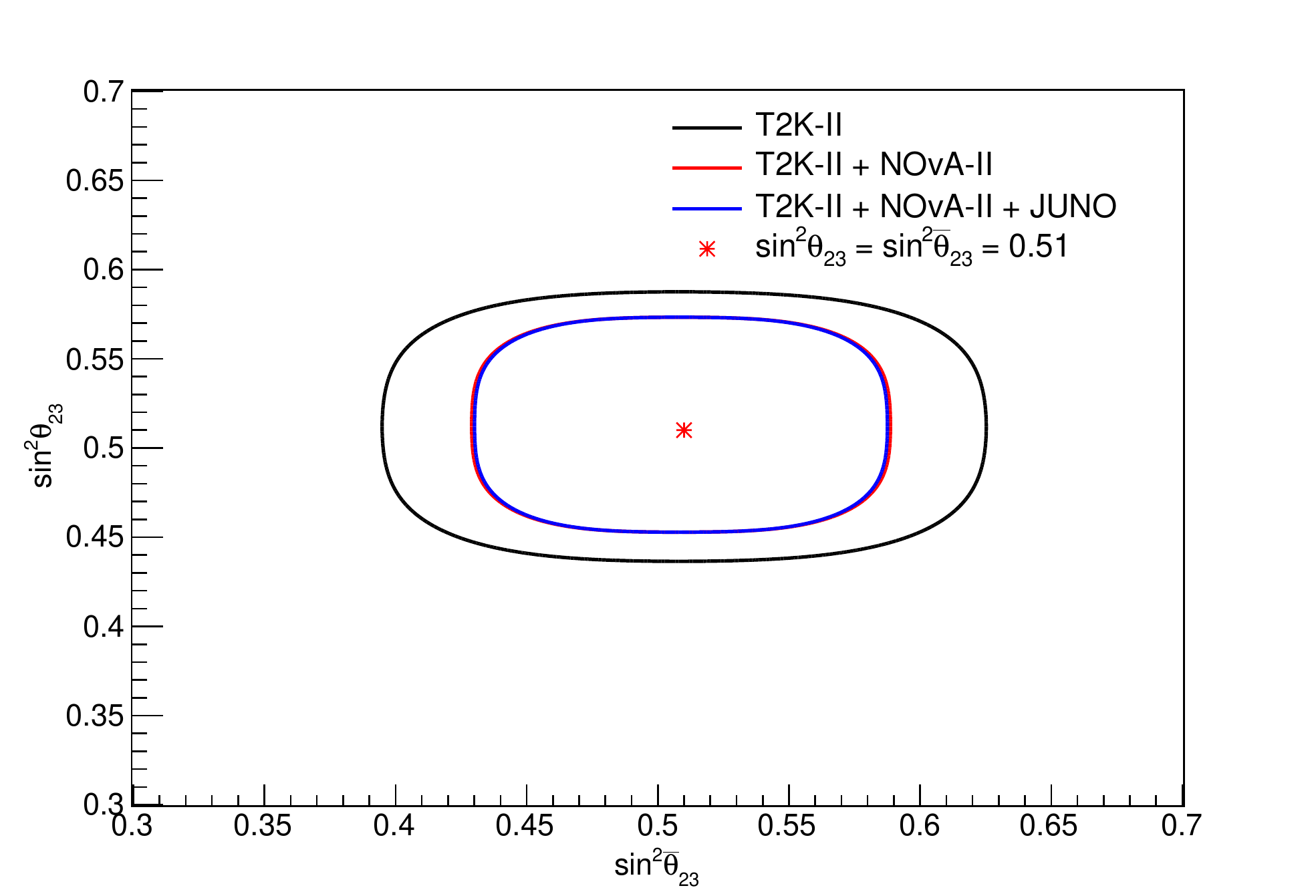}
\caption{\label{fig:sensicontour} The $3\sigma$ C. L. regions of \dmatm\ and $\Delta\overline{m}^2_{31}$ (left), \sinsqthetamu\ and \sinsqthetamubar\   (right). The  black, red, and blue lines are for an analysis with T2K-II only, a joint of T2K-II and \nova-II, and a joint of T2K-II, \nova-II, and JUNO, respectively.}
\end{figure*} 

Fig.~\ref{fig:sensicontour} shows $3\sigma$ C. L. allowed regions of pairs of parameters ($\Delta m^2_{31},\ \Delta\overline{m}^2_{31}$) and ($\sin^2\theta_{23},\ \sin^2\overline{\theta}_{23}$) under the assumption that CPT is conserved. Three different analyses are presented: (i) T2K-II only, (ii) a joint of T2K-II and \nova-II, and (iii) a joint of T2K-II, \nova-II, and JUNO. It is expected that a joint analysis of T2K-II and \nova-II improves significantly the precision of four involved $(\Delta m^2_{31},\ \Delta\overline{m}^2_{31},\ \sin^2\theta_{23},\ \sin^2\overline{\theta}_{23})$ parameters while JUNO mainly contribute to the precision of $\Delta\overline{m}^2_{31}$. 

\begin{figure*}
\centering
\includegraphics[width=0.45\textwidth]{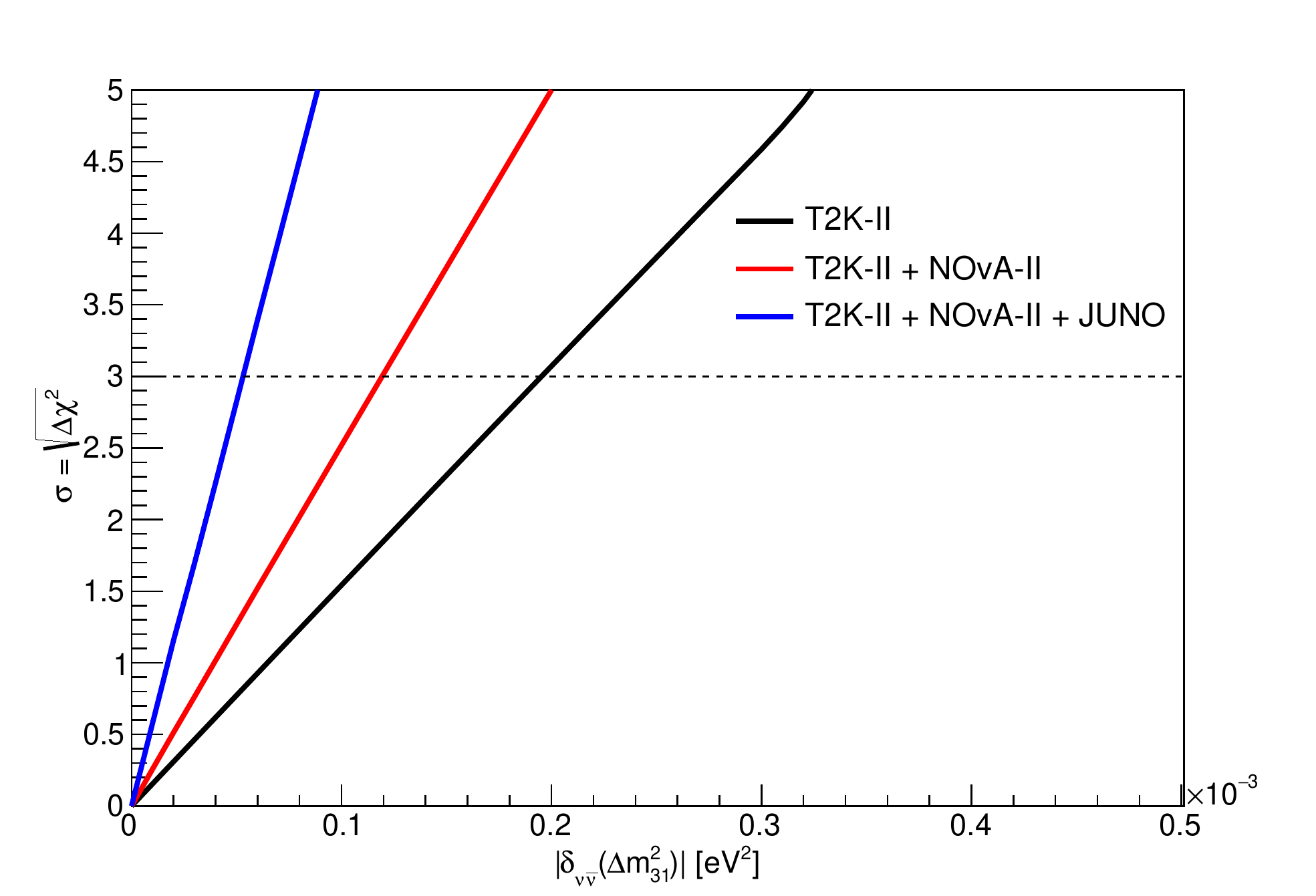}
\includegraphics[width=0.45\textwidth]{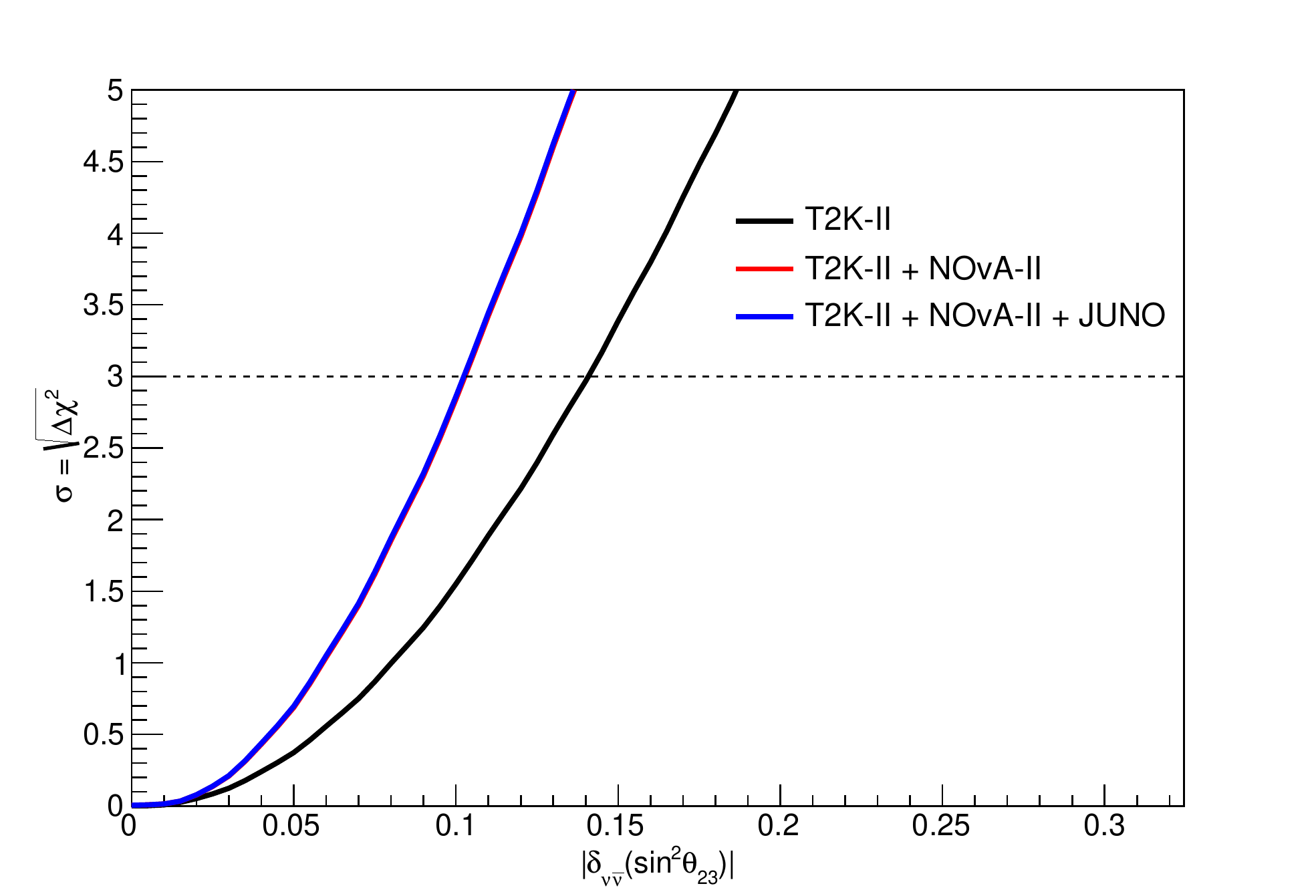}
\caption{\label{fig:sensi1D} The bounds on possible CPT violation manifested in the asymmetries of the mass-squared splittings $|\delta_{\nu\overline{\nu}}(\Delta m^2_{31})|$ (left) and of the leptonic mixing angles $|\delta_{\nu \overline{\nu}}(\sin^2\theta_{23})|$ (right). The black, red, and blue lines correspond to an analysis with T2K-II only, a joint of T2K-II and \nova-II, and a joint of T2K-II, \nova-II, and JUNO, respectively.}
\end{figure*}

To answer for the question about the allowed parameter magnitudes in the mass-squared difference \dcptdm\ and the leptonic mixing angle \dcpttheta, projections of $\Delta \chi^2$ profiles on these two variables are constructed and depicted in Fig.~\ref{fig:sensi1D}. The upper limits of these two CPT-sensitive variables at 3$\sigma$ C. L. are extracted and summarized in Table~\ref{tab:cptdm31h23}. With total exposure of $10\times10^{21}$ POT, T2K-II alone can set more stringent limits on the CPT violation search, if it will be not found, both with atmospheric mass-squared splitting $|\delta_{\nu \overline{\nu}}(\Delta{m}^2_{31})| \leq 2.0\times 10^{-4}~\text{eV}^2$ and leptonic mixing angles $\delta_{\nu \overline{\nu}}(\sin^2\theta_{23})\leq 0.14$, than the combined data of current neutrino experiments. By adding \nova-II, the 3$\sigma$ C. L. limit on $|\delta_{\nu \overline{\nu}}(\sin^2\theta_{23})|$ for CPT violation is reduced to 0.10, a 47\% improvement over the current limit. Meanwhile, if no evidence of CPT violation is found, the potential bound on $|\delta_{\nu\overline{\nu}}(\Delta m^2_{31})|$ at 3$\sigma$ C. L. will be expected to be $5.3\times10^{-5}~\text{eV}^2$ for the combined analysis of the three experiments. This prospective bound on the possible CPT violation search is slightly better than the DUNE sensitivity~\cite{Barenboim2018}, $|\delta_{\nu\overline{\nu}}(\Delta m^2_{31})| < 8.1\times 10^{-5}~\text{eV}^2 ~ \text{at} ~ 3\sigma ~ \text{C. L.}$ 

\begin{table}
\begin{ruledtabular}
 \begin{tabular}{l|c|c}
   &  \multicolumn{2}{c}{3$\sigma$ C. L. upper limits}\\ \hline
 Experiments & $|\delta_{\nu\overline{\nu}}(\Delta m^2_{31})| $ & $|\delta_{\nu\overline{\nu}}(\sin^2\theta_{23})| $ \\ \hline
    T2K-II & $2.0 \times 10^{-4}~ \text{eV}^2$ & 0.14\\
    T2K-II+\nova-II & $1.2 \times 10^{-4}~ \text{eV}^2$ & 0.10\\
    T2K-II+\nova-II+JUNO & $5.3 \times 10^{-5}~ \text{eV}^2$ & 0.10
   \end{tabular}
  \end{ruledtabular}
  \caption{\label{tab:cptdm31h23} The bounds on CPT violation with atmospheric mass-squared difference and mixing angle at $3\sigma$ C. L. for three analyses: T2K-II only, a joint of T2K-II and \nova-II, a joint of T2K-II, \nova-II, and JUNO. }
\end{table}

\section{\label{sec:sensiCPT} Significance of CPT exclusion: dependence and projection}

\begin{figure*}
\includegraphics[width=0.45\textwidth]{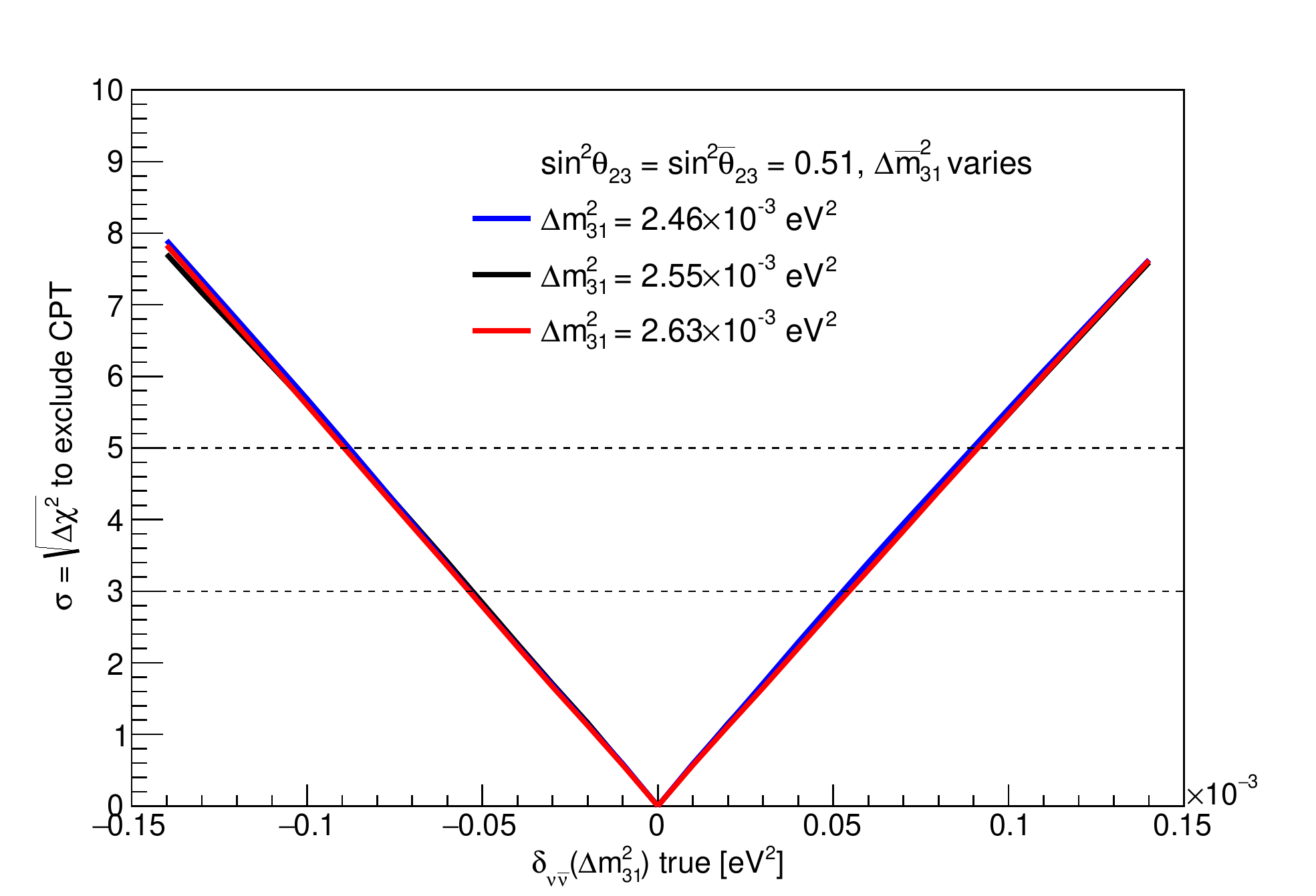}
\includegraphics[width=0.45\textwidth]{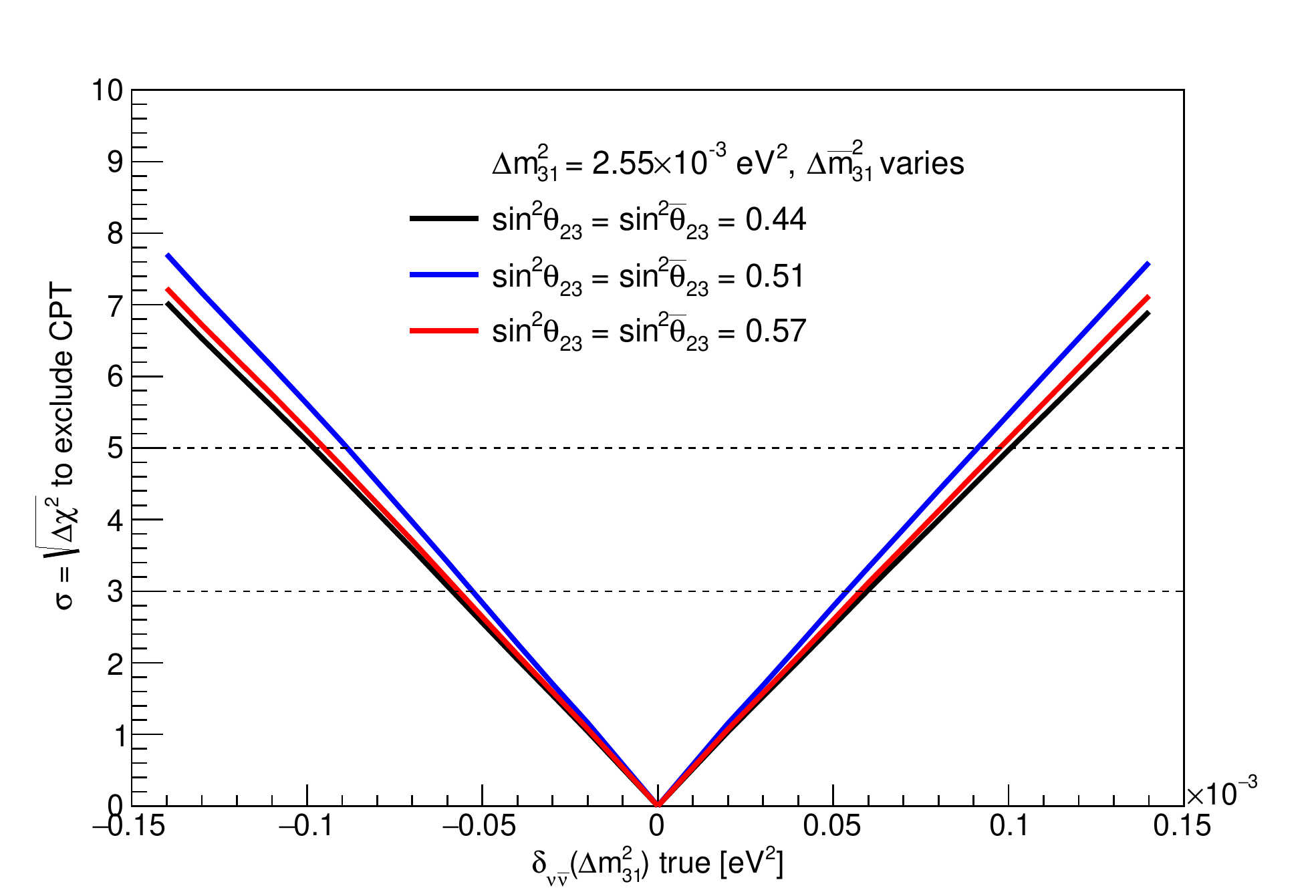}
\caption{\label{fig:dmdepend} Statistical significance to exclude CPT is computed as a function of \textit{true} \dcptdm\ under various scenarios of the involved parameters. The left plot is when \dmatm\ is examined at three different \textit{true} values, while  $\sin^2\theta_{23} =\sin^2\overline{\theta}_{23}=0.51 $ is assumed to be \textit{true}. The right plot presents the CPT sensitivity of \dcptdm\ at different \textit{true} values of  \sinsqthetamu\ and \sinsqthetamubar\   while $\Delta m^2_{31} = 2.55 \times 10^{-3} \text{eV}^2$ is assumed to be \textit{true}.}
\end{figure*}

Apparently if the analyses with real data shows the asymmetries of $|\delta_{\nu\overline{\nu}}(\Delta m^2_{31})| $ or $|\delta_{\nu\overline{\nu}}(\sin^2\theta_{23})|$ larger than the corresponding upper limits presented in Table~\ref{tab:cptdm31h23}, it would imply the CPT violation in the lepton sector. However, one raised question is whether these anticipated limits are affected by the \textit{true} values of the underlying parameters, which can fluctuate from the current best-fit values. To investigate this issue, we performed the full joint analysis of T2K-II, \nova-II, and JUNO under various assumptions of the involved parameters. In particular, for the potential effect on $\delta_{\nu\overline{\nu}}(\Delta m^2_{31})$, we examine the CPT sensitivity at three points ($2.46 \times 10^{-3},~ 2.55\times 10^{-3},~ 2.63 \times 10^{-3} \text{eV}^2$) of \dmatm, taken as the T2K best-fit and $\pm 1\sigma$ shifted values, in combination with a variation of \dmbaratm\ such that $|\delta_{\nu\overline{\nu}}(\Delta m^2_{31})| <0.15\times 10^{-3} \text{eV}^2$. In this case of study, $\sin^2\theta_{23} = \sin^2\overline{\theta}_{23}= 0.51$ is assumed to be \textit{true}. In addition, we check the sensitivities of CPT violation on the $\delta_{\nu\overline{\nu}}(\Delta m^2_{31})$ parameter at three shared values (0.44, 0.51, 0.57) of $(\sin^2\theta_{23},\ \sin^2\overline{\theta}_{23})$. For each case, the statistical significance to exclude the corresponding form of the CPT invariance is extracted as function of \dcptdm\ and the results are shown in Fig.~\ref{fig:dmdepend}.
\noindent It is observed that the CPT violation sensitivity manifested on the \dcptdm\ parameter depend marginally on the central value of \dmatm\ and $\Delta \overline{m}^2_{31}$ in the current allowed range of this parameter. Also the dependence of the \dcptdm\ sensitivity on the \textit{true} value of the mixing parameter $(\sin^2\theta_{23},\ \sin^2\overline{\theta}_{23})$ is relatively small. Apparently, due to the octant degeneracy of  $(\sin^2\theta_{23},\ \sin^2\overline{\theta}_{23})$ presented in the \textit{disappearance} probabilities of muon (anti-)neutrinos, the significance of the CPT test is slightly worse than the case where $(\sin^2\theta_{23},\ \sin^2\overline{\theta}_{23})$  is exactly equal or near the maximal mixing. The lower limit of \textit{true} \dcptdm\ magnitude to exclude the CPT at 3$\sigma$ C. L. or higher significance is presented in Table~\ref{tab:cptsensidm31}.
 \begin{table}
     \centering
\begin{ruledtabular}
 \begin{tabular}{c|c|c|c}
\multirow{2}{*}{\dmatm\ [$\text{eV}^{2}$]} &  \multicolumn{3}{c}{Shared values of $\sin^2\theta_{23},\ \sin^2\overline{\theta}_{23}$}\\ \cline{2-4}
 & 0.44 & 0.51 & 0.57 \\
\hline
    $2.46\times 10^{-3}$ & $5.96\times 10^{-5}$ & $5.36\times 10^{-5}$ & $5.80\times 10^{-5}$ \\
    $2.55\times 10^{-3}$ & $5.95\times 10^{-5}$ & $5.39\times 10^{-5}$ & $5.77\times 10^{-5}$ \\ 
    $2.63\times 10^{-3}$ & $5.99\times 10^{-5}$ & $5.46\times 10^{-5}$ & $5.79\times 10^{-5}$ \\
    \cline{2-4}
   &\multicolumn{3}{c}{ \dcptdm\ limit to exclude CPT at  3 $\sigma$ C. L.}\\
   \end{tabular}
  \end{ruledtabular}
  \caption{\label{tab:cptsensidm31} Lower limits for the \textit{true} $|\delta_{\nu\overline{\nu}}(\Delta m^2_{31})|$ magnitude to exclude CPT at 3$\sigma$ C. L. are computed at different \textit{true} values of the involved parameters.}
\end{table}
 \noindent We find that if the deviation of \dcptdm\ from zero is greater than $6.0\times 10^{-5} \text{eV}^2$ the CPT invariance will be excluded at 3$\sigma$ C. L. for almost the entire currently-allowed range of the involved parameters. The range of possible \dcptdm\  asymmetry to be explored significantly is slightly extended ($[5.36,5.46]\times 10^{-5} \text{eV}^2$) if the mixing angle is near the maximal mixing. Due to the aforementioned octant degeneracy of the (anti-)neutrino oscillation probabilities in the \textit{disappearance} samples, the deviation of \dcptdm\ from zero must be moderately greater ($[5.77,5.99]\times 10^{-5} \text{eV}^2$) for attaining a same level of significance to exclude the CPT invariance. To see how impressive the improvement in the CPT test sensitivity from this three-experiment combined analysis is, we project the statistical significance from the current measurements. As summarized in the Table~\ref{tab:atmos_paras_experiments}, the difference in mass-squared splitting at the best-fit values of $(\Delta m^2_{31},\ \Delta \overline{m}^2_{31})$ measured by T2K~\cite{abe2021t2k} is $|\delta_{\nu\overline{\nu}}(\Delta m^2_{31})| = 3\times10^{-5} \text{eV}^2$, well consistent within 1$\sigma$ uncertainty of $20\times10^{-5} \text{eV}^2$. However, if this asymmetry persists as the \textit{true}, it will correspond to $1.7\sigma$~C.~L. exclusion of CPT conservation by the combined analysis of T2K-II, \nova-II, and JUNO. If the level of asymmetrical \dcptdm\ in the neutrino and anti-neutrino best-fit values of \nova\ and MINOS(+), which is $7.0\times 10^{-5}\ \text{eV}^2$, are assumed to be persisted as the \textit{true}, the synergy of the three experiments can exclude CPT conservation at $4\sigma$ C. L. 

\begin{figure*}
\includegraphics[width=0.45\textwidth]{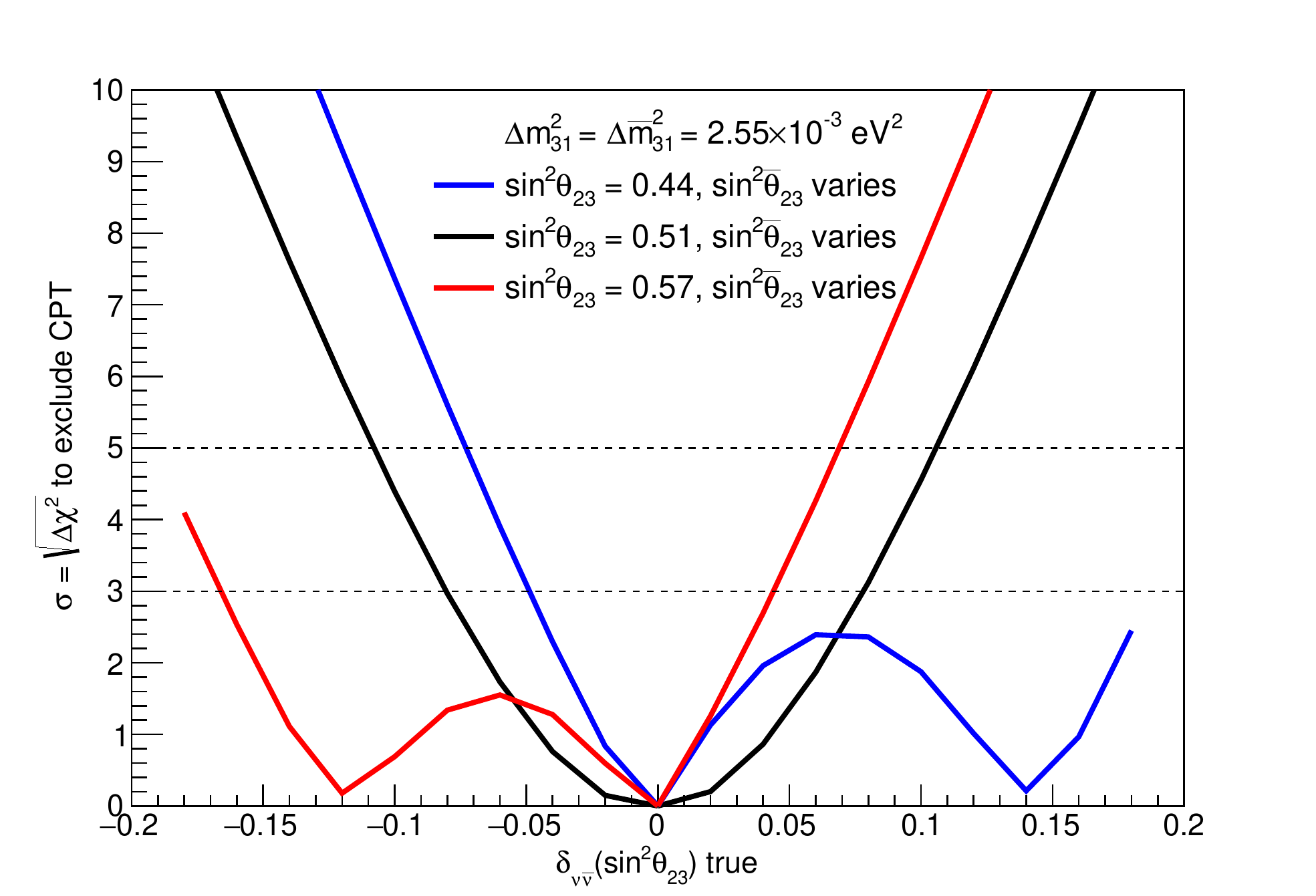}
\includegraphics[width=0.45\textwidth]{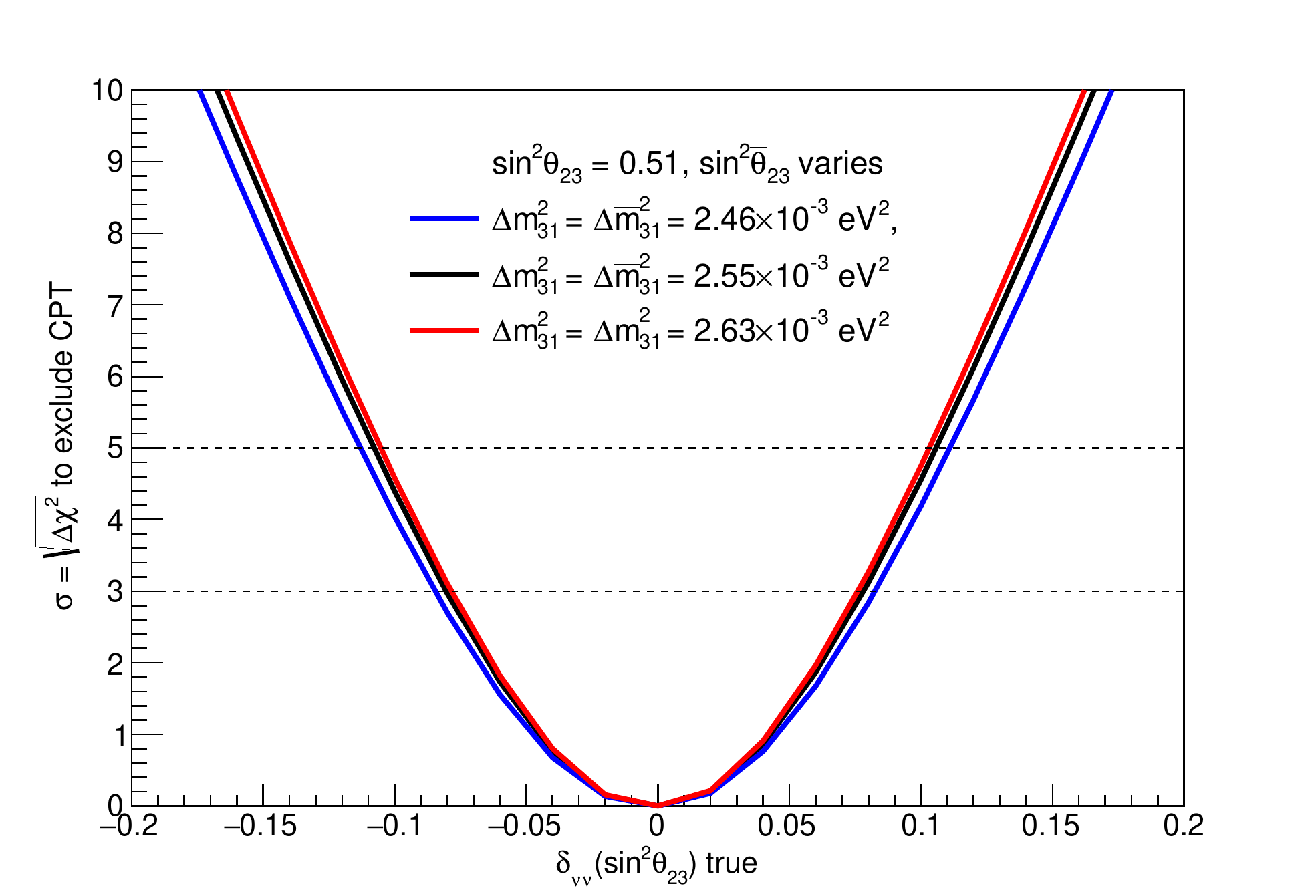}
\caption{\label{fig:th23depend} Statistical significance to exclude CPT is computed as function of \textit{true} \dcpttheta\ under various scenarios of the involved parameters. The left is when \sinsqthetamu\ is examined at three different \textit{true} values while 
$\Delta m^2_{31} = \Delta \overline{m}^2_{31}= 2.55 \times 10^{-3} \text{eV}^2$
is assumed. The right presents the CPT sensitivity of \dcpttheta\ at different \textit{true} values of  \dmatm\ and \dmbaratm\ while $\sin^2\theta_{23}=0.51$ is assumed to be \textit{true}.}
\end{figure*}

\begin{figure}
\includegraphics[width=0.45\textwidth]{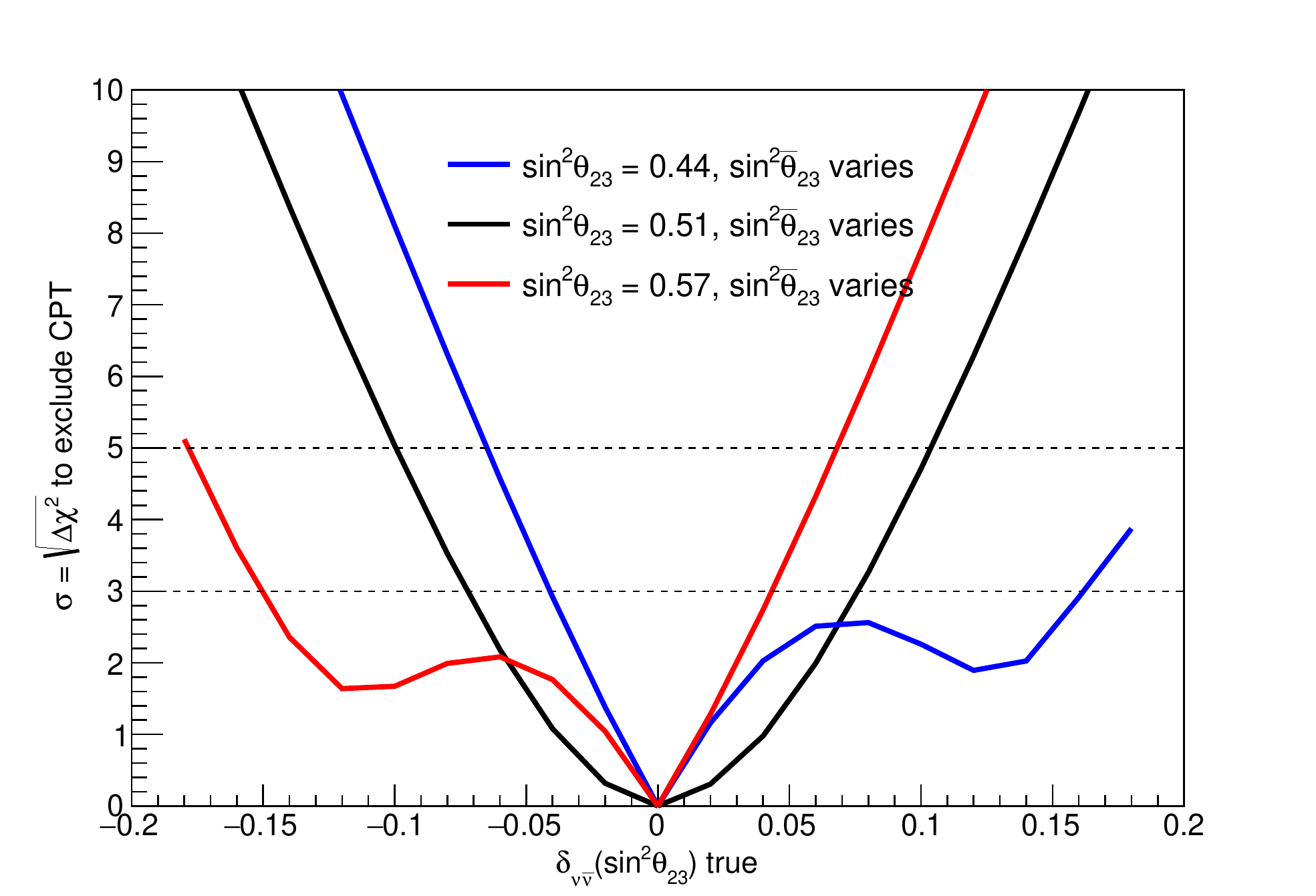}
\caption{\label{fig:th23allsample} Statistical significance to exclude CPT is computed as a function of \textit{true} \dcpttheta\ under various scenarios of the involved parameters. Both muon (anti-)neutrino \textit{disappearance} samples and electron (anti-)neutrino \textit{appearance} samples from T2K-II and \nova-II are used.  The sensitivity is examined at three different \textit{true} values of \sinsqthetamu\ values while $\Delta m^2_{31} = \Delta \overline{m}^2_{31}= 2.55 \times 10^{-3} \text{eV}^2$
is assumed to be \textit{true}.}
\end{figure}

Regarding the sensitivity of \dcpttheta\ on the CPT test, we examine and find that their dependence on the fluctuation of the $(\Delta m^2_{31},\ \Delta \overline{m}^2_{31})$ parameters is relatively small while the dependence on the \textit{true} value of (\sinsqthetamu,\ \sinsqthetamubar) is significant, as shown in Fig.~\ref{fig:th23depend}. When the \textit{true} value of \sinsqthetamu\ belongs to an octant, there exists a degenerated solution in the other octant. For example, when \sinsqthetamu=0.44, the \textit{extrinsic} CPT-invariant solution of \sinsqthetamubar=0.58 (along with the \textit{genuine} solution of \sinsqthetamubar=0.44). Similar behavior is observed when \sinsqthetamu\ values in the higher octant. The behavior is well-understood due to the dependence of muon (anti-) neutrino \textit{disappearance} probabilities on the \sinsqtwothetamu\ (\sinsqtwothetamubar) rather than \sinsqthetamu\ (\sinsqthetamubar).
\noindent As summarized in Table~\ref{tab:cptsensith23}, to attain the same significance level to exclude the CPT, compared to the maximal case \sinsqthetamu=0.51, the magnitude of \textit{true} \dcpttheta\ asymmetry in the non-maximal cases (\sinsqthetamu=0.44 and \sinsqthetamu=0.57) is required to be larger or smaller depending on whether the \thetamu\ and \thetamubar\ belong to the different or same octants, respectively. In particular, for \sinsqthetamu=0.51 as indicated by both T2K~\cite{abe2021t2k} and \nova\ \cite{nova20192d118E20POT}, the magnitude of \dcpttheta\ asymmetry must be between [0.076, 0.084] to be discovered with 3$\sigma$ C. L. T2K (\nova) measured \dcpttheta=0.08 (0.10) respectively, and if it remains as \textit{true} the CPT invariance will be excluded at 3$\sigma$ or higher C. L. If \thetamu\ and \thetamubar\ are in the same octant and relatively far off from the maximal values, the deviation of \dcpttheta\ from zero must be greater than 0.051 in order to rule out CPT invariance at 3$\sigma$ C. L.. If \thetamu\ and \thetamubar\ are in different octants, \thetamu\ in lower octant and \thetamubar\ in higher octant or vice versa, the magnitude of \dcpttheta\ must be significantly higher, varying in the (0.165,0.190) range, to exclude CPT at the same 3$\sigma$ statistical significance.
\begin{table}
     \centering
\begin{ruledtabular}
 \begin{tabular}{c|c|c|c}
  \multirow{2}{*}{ \sinsqthetamu} &  \multicolumn{3}{c}{Shared values of \dmatm, $\Delta\overline{m}^2_{31}$~[$\text{eV}^2$]}\\ \cline{2-4}
 & $2.46\times 10^{-3}$ & $2.55\times 10^{-3}$ & $2.63\times 10^{-3}$ \\
  \hline
    0.44  & -0.051 (+0.190) & -0.049 (+0.187) & -0.048 (+0.186) \\
    0.51 & -0.084 (+0.082) & -0.080 (+0.078) & -0.078 (+0.076) \\ 
    0.57 & -0.169 (+0.047) & -0.166 (+0.044) & -0.165 (+0.043) \\
        \cline{2-4}
   &\multicolumn{3}{c}{ $\delta_{\nu\overline{\nu}}(\sin^2 \theta_{23})$ limit to exclude CPT at  3 $\sigma$ C. L.}\\
   \end{tabular}
  \end{ruledtabular}
  \caption{\label{tab:cptsensith23} 
  Lower limits for the \textit{true} $|\delta_{\nu\overline{\nu}}(\sin^2\theta_{23})|$ deviation from zero to exclude CPT at 3$\sigma$ C. L. are computed at different \textit{true} values of involved parameters. The -(+) signs in each cell correspond to the negative (positive) value of \dcpttheta. }
\end{table}
The sensitivity to detect CPT violation via the \dcpttheta\ asymmetry is not good due to the aforementioned octant degeneracy in the muon (anti-) neutrino \textit{disappearance} samples. The sensitivity can be improved by adding the electron (anti-)neutrino \textit{appearance} samples from the A-LBL experiments. Fig.~\ref{fig:th23allsample} shows the sensitivity of \dcpttheta\ on the CPT exclusion with a combination of both \textit{disappearance} and \textit{appearance} samples.
\noindent It is observed that by adding the electron (anti-)neutrino \textit{appearance} samples, the statistical significance to exclude the \textit{extrinsic} CPT-invariant solution is enhanced notably. Consequently, the sensitivity of \dcpttheta\ to the CPT violation has improved. However, one must consider carefully when adding the electron (anti-)neutrino \textit{appearance} samples. The reason is that the probabilities of \nue(\nueb) from \numu(\numub) depend not only on \thetamu(\thetamubar) but also on two known unknowns, \textit{CP-}violating phase and mass ordering, which will complicate the interpretation of the experimental observation.  

\section{\label{sec:fin}Conclusion}
In the paper, we presented the potential of timely combined analysis of the two on-going accelerator-based long-baseline experiments T2K-II, \nova-II and a reactor-based medium-baseline JUNO experiment in testing CPT symmetry via the measurable asymmetry in the oscillation parameters of neutrinos and anti-neutrinos. The analysis is expected to happen by 2028, before the operational start of the next generation of accelerator-based long-baseline neutrino experiments, DUNE~\cite{DUNE:2020lwj} and Hyper-Kamiokande~\cite{Abe:2018uyc}. In particular, we focus on the asymmetries in the mass-square splitting \dcptdm\ and in the leptonic mixing angle \dcpttheta. The synergy of these three experiment will plausibly establish an unprecedented bound of \dcptdm\ to about $5.3\times 10^{-5} \text{eV}^2$ at 3$\sigma$ C. L. in case the CPT symmetry is conserved to be sensitive. This bound extends substantially the current bound of $2.5\times 10^{-4} \text{eV}^2$ derived from the global neutrino data analysis. It is noteworthy to stress that this bound of \dcptdm\ on the possible CPT violation is marginally dependent on the \textit{true} values of the involved parameters, especially the ambiguity of the \thetamu(\thetamubar) values. The improvement of CPT sensitivity is very encouraging since if the difference between the best-fit values of \dmatm\ and \dmbaratm currently measured by \nova\ and MINOS(+) persists as the \textit{true}, the statistical significance to exclude the CPT is about 4$\sigma$ C. L.. For the testable asymmetry in the leptonic mixing angle \dcpttheta, the statistical significance of the CPT sensitivity depends strongly on their own \textit{genuine} values, which rooted from the parameter degeneracy in the muon (anti-)neutrino \textit{disappearance} probabilities. In the case of CPT conservation, if the neutrino mixing angle \thetamu\ is close to the maximal, as indicated by both T2K and \nova\ current measurements, the combined analysis of the three experiments will potentially establish a limit of \dcpttheta=0.10, compared to the current bound of 0.19 attained from the global neutrino data analysis. Interestingly, if the difference in the best-fit values of \sinsqthetamu\ and \sinsqthetamubar\ measured recently by both T2K and \nova\ persist as the \textit{true}, the combined analysis of the two with their final data samples will indicate CPT violation with 3$\sigma$ C. L. or higher.

Finally, it is important to emphasize that one cannot claim a CPT violation simply by observing sizable \dcptdm\ or \dcpttheta\ asymmetries because some non-standard interactions, such as those discussed in Ref.~\cite{Ge:2019tdi}, can mimic the effect. In any case, investigating the potential differences in the parameters governing neutrino and anti-neutrino oscillations is critical to revealing the new physics.
\vspace{1cm}
\section*{Acknowledgement}
T. V. Ngoc was funded by Vingroup JSC and supported by the Master, PhD Scholarship Programme of Vingroup Innovation Foundation (VINIF), Institute of Big Data, code VINIF.2021.TS.069.

\bibliographystyle{jhep} 
\bibliography{apssamp}

\providecommand{\noopsort}[1]{}\providecommand{\singleletter}[1]{#1}%

\providecommand{\href}[2]{#2}\begingroup\raggedright\begin{thebibliography}{10}

\bibitem{Luders1954}
G.~Luders, \emph{{On the Equivalence of Invariance under Time Reversal and
  under Particle-Antiparticle Conjugation for Relativistic Field Theories}},
  {\emph{Kong. Dan. Vid. Sel. Mat. Fys. Med.} {\bfseries 28N5} (1954) 1}.

\bibitem{1955nbdpbookPauli}
W.~{Pauli}, L.~{Rosenfeld} and V.~{Weisskopf}, \emph{{Niels Bohr and the
  Development of Physics}}. {Pergamon Press}, 1955.

\bibitem{Jost1957}
R.~{Jost}, \emph{{Eine Bemerkung zum CTP theorem}}, {\emph{Helv. Phys. Acta}
  {\bfseries 30} (1957) }.

\bibitem{Schwinger1958}
J.~S. Schwinger, \emph{{Spin, statistics, and the TCP theorem}},
  \href{https://doi.org/10.1073/pnas.44.2.223}{\emph{Proc. Nat. Acad. Sci.}
  {\bfseries 44} (1958) 223}.

\bibitem{johnbell}
A.~{Whitaker}, \emph{{ John Stewart Bell and Twentieth-Century Physics}}.
  {Oxford University Press}, 2016.

\bibitem{Kosteleck_2011}
V.~A. Kosteleck{\'{y} } and N.~Russell, \emph{Data tables for lorentz and cpt
  violation}, \href{https://doi.org/10.1103/revmodphys.83.11}{\emph{Reviews of
  Modern Physics} {\bfseries 83} (2011) 11}.

\bibitem{CPTsummary2022}
C.~A. Argüelles, G.~Barenboim, M.~Bustamante, P.~Coloma, P.~B. Denton,
  I.~Esteban et~al., \emph{Snowmass white paper: Beyond the standard model
  effects on neutrino flavor},  2022.
\newblock 10.48550/ARXIV.2203.10811.

\bibitem{pdg2020}
{\scshape Particle Data Group} collaboration, \emph{{Review of Particle
  Physics}}, \href{https://doi.org/10.1093/ptep/ptaa104}{\emph{PTEP} {\bfseries
  2020} (2020) 083C01}.

\bibitem{Murayama_2004}
H.~Murayama, \emph{{CPT tests: Kaon versus neutrinos}},
  \href{https://doi.org/10.1016/j.physletb.2004.06.106}{\emph{Phys. Lett. B}
  {\bfseries 597} (2004) 73}
  [\href{https://arxiv.org/abs/hep-ph/0307127}{{\ttfamily hep-ph/0307127}}].

\bibitem{Barger_2000}
V.~D. Barger, S.~Pakvasa, T.~J. Weiler and K.~Whisnant, \emph{{CPT odd
  resonances in neutrino oscillations}},
  \href{https://doi.org/10.1103/PhysRevLett.85.5055}{\emph{Phys. Rev. Lett.}
  {\bfseries 85} (2000) 5055}
  [\href{https://arxiv.org/abs/hep-ph/0005197}{{\ttfamily hep-ph/0005197}}].

\bibitem{Cao:2021ptu}
S.~Cao, N.~T. Hong~Van, T.~V. Ngoc and P.~T. Quyen, \emph{{Neutrino Mass
  Spectrum: Present Indication and Future Prospect}},
  \href{https://arxiv.org/abs/2111.11644}{{\ttfamily 2111.11644}}.

\bibitem{Barenboim:2002hx}
G.~Barenboim, J.~F. Beacom, L.~Borissov and B.~Kayser, \emph{{CPT Violation and
  the Nature of Neutrinos}},
  \href{https://doi.org/10.1016/S0370-2693(02)01947-0}{\emph{Phys. Lett. B}
  {\bfseries 537} (2002) 227}
  [\href{https://arxiv.org/abs/hep-ph/0203261}{{\ttfamily hep-ph/0203261}}].

\bibitem{maki1962remarks}
Z.~Maki, M.~Nakagawa and S.~Sakata, \emph{Remarks on the unified model of
  elementary particles}, \href{https://doi.org/10.1143/PTP.28.870}{\emph{Prog.
  Theor. Phys.} {\bfseries 28} (1962) 870}.

\bibitem{pontecorvo1958mesonium}
B.~Pontecorvo, \emph{Mesonium and antimesonium}, {\emph{JETP} {\bfseries 6}
  (1958) 429}.

\bibitem{RevModPhys.88.030501}
T.~Kajita, \emph{Nobel lecture: Discovery of atmospheric neutrino
  oscillations}, \href{https://doi.org/10.1103/RevModPhys.88.030501}{\emph{Rev.
  Mod. Phys.} {\bfseries 88} (2016) 030501}.

\bibitem{RevModPhys.88.030502}
A.~B. McDonald, \emph{Nobel lecture: The sudbury neutrino observatory:
  Observation of flavor change for solar neutrinos},
  \href{https://doi.org/10.1103/RevModPhys.88.030502}{\emph{Rev. Mod. Phys.}
  {\bfseries 88} (2016) 030502}.

\bibitem{T2Knature2020}
{\scshape T2K} collaboration, \emph{{Constraint on the matter–antimatter
  symmetry-violating phase in neutrino oscillations}},
  \href{https://doi.org/https://doi.org/10.1038/s41586-020-2177-0}{\emph{Nature}
  {\bfseries 580} (2020) 339}.

\bibitem{Barenboim2020}
G.~Barenboim, C.~A. Ternes and M.~T{\'{o}}rtola, \emph{{CPT} and {CP}, an
  entangled couple},
  \href{https://doi.org/10.1007/jhep07(2020)155}{\emph{Journal of High Energy
  Physics} {\bfseries 2020} (2020) }.

\bibitem{Barenboim_2001}
M.~Ba{\~{n}}uls, G.~Barenboim and J.~Bernab{\'{e}}u, \emph{Medium effects for
  terrestrial and atmospheric neutrino oscillations},
  \href{https://doi.org/10.1016/s0370-2693(01)00723-7}{\emph{Physics Letters B}
  {\bfseries 513} (2001) 391}.

\bibitem{Xing_2002}
Z.~zhong Xing, \emph{Fake {CPT} violation in disappearance neutrino
  oscillations},
  \href{https://doi.org/10.1088/0954-3899/28/2/402}{\emph{Journal of Physics G:
  Nuclear and Particle Physics} {\bfseries 28} (2002) B7}.

\bibitem{Bernabeu_2002}
J.~Bernab{\'{e}}u, S.~Palomares-Ruiz, A.~P{\'{e}}rez and S.~Petcov, \emph{The
  earth mantle-core effect in charge-asymmetries for atmospheric neutrino
  oscillations},
  \href{https://doi.org/10.1016/s0370-2693(02)01358-8}{\emph{Physics Letters B}
  {\bfseries 531} (2002) 90}.

\bibitem{Jacobson_2004}
M.~Jacobson and T.~Ohlsson, \emph{{Extrinsic CPT violation in neutrino
  oscillations in matter}},
  \href{https://doi.org/10.1103/PhysRevD.69.013003}{\emph{Phys. Rev. D}
  {\bfseries 69} (2004) 013003}
  [\href{https://arxiv.org/abs/hep-ph/0305064}{{\ttfamily hep-ph/0305064}}].

\bibitem{OHLSSON2015482}
T.~Ohlsson and S.~Zhou, \emph{{Extrinsic and intrinsic CPT asymmetries in
  neutrino oscillations}},
  \href{https://doi.org/https://doi.org/10.1016/j.nuclphysb.2015.02.015}{\emph{Nuclear
  Physics B} {\bfseries 893} (2015) 482 }.

\bibitem{abe2021t2k}
{\scshape T2K} collaboration, \emph{T2k measurements of muon neutrino and
  antineutrino disappearance using
  $3.13\ifmmode\times\else\texttimes\fi{}{10}^{21}$ protons on target},
  \href{https://doi.org/10.1103/PhysRevD.103.L011101}{\emph{Phys. Rev. D}
  {\bfseries 103} (2021) L011101}.

\bibitem{deSalas2020}
P.~F. de~Salas, D.~V. Forero, S.~Gariazzo, P.~Martinez-Mirave, O.~Mena, C.~A.
  Ternes et~al., \emph{{2020 global reassessment of the neutrino oscillation
  picture}}, \href{https://doi.org/10.1007/JHEP02(2021)071}{\emph{JHEP}
  {\bfseries 02} (2021) 071}
  [\href{https://arxiv.org/abs/2006.11237}{{\ttfamily 2006.11237}}].

\bibitem{SonCao2020}
S.~Cao, A.~Nath, T.~V. Ngoc, P.~T. Quyen, N.~T. Hong~Van and N.~K. Francis,
  \emph{{Physics potential of the combined sensitivity of T2K-II, NO$\nu$A
  extension, and JUNO}},
  \href{https://doi.org/10.1103/PhysRevD.103.112010}{\emph{Phys. Rev. D}
  {\bfseries 103} (2021) 112010}
  [\href{https://arxiv.org/abs/2009.08585}{{\ttfamily 2009.08585}}].

\bibitem{Adamson2013}
P.~Adamson et~al., \emph{Measurement of neutrino and antineutrino oscillations
  using beam and atmospheric data in {MINOS}},
  \href{https://doi.org/10.1103/physrevlett.110.251801}{\emph{Physical Review
  Letters} {\bfseries 110} (2013) }.

\bibitem{minos2020}
{\scshape $\mathrm{MINOS}+$} collaboration, \emph{Precision constraints for
  three-flavor neutrino oscillations from the full $\mathrm{MINOS}+$ and minos
  dataset}, \href{https://doi.org/10.1103/PhysRevLett.125.131802}{\emph{Phys.
  Rev. Lett.} {\bfseries 125} (2020) 131802}.

\bibitem{nova20192d118E20POT}
D.~P. Mendez~Mendez, \emph{Comparison of oscillation parameters measured from
  $\nu_\mu$ and $\bar{\nu}_\mu$ disappearance in the NOvA experiment}, Ph.D.
  thesis, University of Sussex, 2019.

\bibitem{dayabay2022}
K.~B. Luk, \emph{{Reactor Neutrino: Latest Results from Daya Bay}},  June,
  2022.
\newblock 10.5281/zenodo.6683712.

\bibitem{Huber:2004ka}
{Huber, Patrick and Lindner, M., and Winter, W.}, \emph{Simulation of
  long-baseline neutrino oscillation experiments with {GLoBES (General Long
  Baseline Experiment Simulator)}},
  \href{https://doi.org/10.1016/j.cpc.2005.01.003}{\emph{Comput. Phys. Commun.}
  {\bfseries 167} (2005) 195} [\href{https://arxiv.org/abs/0407333}{{\ttfamily
  0407333}}].

\bibitem{huber2007new}
{Huber, Patrick and Kopp, Joachim and Lindner, Manfred and Rolinec, Mark, and
  Winter, Walter}, \emph{New features in the simulation of neutrino oscillation
  experiments with {GLoBES 3.0}},
  \href{https://doi.org/10.1016/j.cpc.2007.05.004}{\emph{Comput. Phys. Commun.}
  {\bfseries 177} (2007) 432}
  [\href{https://arxiv.org/abs/hep-ph/0701187}{{\ttfamily hep-ph/0701187}}].

\bibitem{Barger1980}
V.~Barger, K.~Whisnant, S.~Pakvasa and R.~J.~N. Phillips, \emph{Matter effects
  on three-neutrino oscillations},
  \href{https://doi.org/10.1103/PhysRevD.22.2718}{\emph{Phys. Rev. D}
  {\bfseries 22} (1980) 2718}.

\bibitem{Abe:2011ks}
{\scshape T2K} collaboration, \emph{The {T2K} experiment},
  \href{https://doi.org/10.1016/j.nima.2011.06.067}{\emph{Nucl. Instrum.
  Methods Phys. Res. Sect. A} {\bfseries 659} (2011) 106}
  [\href{https://arxiv.org/abs/1106.1238}{{\ttfamily 1106.1238}}].

\bibitem{ayres2007nova}
{\scshape NO$\nu$A} collaboration, \emph{{The NOvA Technical Design Report}},
  Tech. Rep. FERMILAB-DESIGN-2007-01, 10, 2007.
\newblock 10.2172/935497.

\bibitem{christophe_bronner_2022_6683821}
C.~Bronner, \emph{{Accelerator Neutrino I\_Recent results from T2K}},  June,
  2022.
\newblock 10.5281/zenodo.6683821.

\bibitem{jeff_hartnell_2022_6683827}
J.~Hartnell, \emph{{Accelerator Neutrino I\_NOvA}},  June, 2022.
\newblock 10.5281/zenodo.6683827.

\bibitem{djurcic2015juno}
{\scshape JUNO} collaboration, \emph{{JUNO Conceptual Design Report}},
  \href{https://arxiv.org/abs/1508.07166}{{\ttfamily 1508.07166}}.

\bibitem{juno_2022}
{JUNO Collaboration}, A.~Abusleme et~al., \emph{{Sub-percent Precision
  Measurement of Neutrino Oscillation Parameters with JUNO}},
  \href{https://arxiv.org/abs/2204.13249}{{\ttfamily 2204.13249}}.

\bibitem{abe2016sensitivity}
{\scshape T2K} collaboration, \emph{Sensitivity of the {T2K} accelerator-based
  neutrino experiment with an extended run to $20\times10^{21}$ {POT}s},
  \href{https://arxiv.org/abs/1607.08004}{{\ttfamily 1607.08004}}.

\bibitem{t2k_collaboration_2021_flux}
{T2K Collaboration}, \emph{Neutrino beam flux prediction 2020},  Nov., 2021.
\newblock 10.5281/zenodo.5734307.

\bibitem{Barenboim2018}
G.~Barenboim, C.~Ternes and M.~Tórtola, \emph{{Neutrinos, DUNE and the world
  best bound on CPT invariance}},
  \href{https://doi.org/https://doi.org/10.1016/j.physletb.2018.03.060}{\emph{Physics
  Letters B} {\bfseries 780} (2018) 631 }.

\bibitem{DUNE:2020lwj}
{\scshape DUNE} collaboration, \emph{{Deep Underground Neutrino Experiment
  (DUNE), Far Detector Technical Design Report, Volume I Introduction to
  DUNE}}, \href{https://doi.org/10.1088/1748-0221/15/08/T08008}{\emph{JINST}
  {\bfseries 15} (2020) T08008}
  [\href{https://arxiv.org/abs/2002.02967}{{\ttfamily 2002.02967}}].

\bibitem{Abe:2018uyc}
{\scshape Hyper-Kamiokande Proto-Collaboration} collaboration,
  \emph{Hyper-{K}amiokande {D}esign {R}eport},
  \href{https://arxiv.org/abs/1805.04163}{{\ttfamily 1805.04163}}.

\bibitem{Ge:2019tdi}
S.-F. Ge and H.~Murayama, \emph{{Apparent CPT Violation in Neutrino Oscillation
  from Dark Non-Standard Interactions}},
  \href{https://arxiv.org/abs/1904.02518}{{\ttfamily 1904.02518}}.

\end{thebibliography}\endgroup
\end{document}